\documentclass[aps,longbibliography,nofootinbib,groupedaddress,twocolumn,showpacs,preprintnumbers,amsmath,amssymb]{revtex4-1}

\usepackage{graphicx}
\usepackage{bm}
\usepackage{gensymb}
\usepackage{textcomp}  

\begin{document}

\preprint{}

\title{Comment on ``Relativistic effects in atom and neutron interferometry and the differences between them''}

\author{Hartmut Lemmel$^{1,2}$}
\email{hartmut@lemmel.at}
\affiliation{$^1$Vienna University of Technology, Atominstitut, 1020 Wien, Austria;
 $^2$Institut Laue Langevin, 38000 Grenoble, France}

\date{\today}

\begin{abstract}

\end{abstract}

\maketitle


Greenberger et al investigated in a recent paper \cite{greenberger2012} whether relativistic effects like redshift and twin paradox could be seen in an atom or neutron interferometer in the gravitational field. They find that the observability of such effects depends on the mechanism of reflection by the mirrors of the interferometer. On its path between beam splitter and mirror the particle accumulates a certain vertical momentum change due to gravity. If this accumulated momentum change is reflected by the mirror, both beam paths accumulate a different amount of proper time and the observed phase shift can be interpreted as a relativistic effect.

In the Kasevich-Chu experiment the atom is reflected by a laser pulse where a certain momentum is transferred onto the atom by photon exchange. The accumulated gravitational momentum change is not affected. When the beam paths come together, the atom has fallen by the same amount that it would have fallen without interferometer and there is no difference in proper time between the beam paths. The observed phase shift has no relativistic origin \cite{schleich2013}.

 In the COW experiment the neutron is reflected by the lattice planes of a Bragg crystal in transmission geometry (symmetric Laue case). The authors state that the complete neutron momentum is reflected including the accumulated gravitational momentum change. They compare the reflection process with the elastic bounce of a ball on a flat surface. This we want to object. Bragg diffraction is an interference phenomenon based on multiple scattering. The process is correctly described by dynamical diffraction (e.g. \cite{batterman1964,rauch-werner}) rather than classical mechanics. It then follows for the symmetric Laue case that the momentum transfer perpendicular to the lattice planes is always $\hbar \vec H$ (cf. e.g. \cite{petrascheck1976,lemmel2007}) with $\vec H$ denoting the reciprocal lattice vector. Its modulus is given by $H=2 \pi/d$ where $d$ denotes the lattice constant. As in the atom case, the accumulated gravitational momentum change is not affected, and both types of interferometers are comparable to lowest order.
 Nevertheless, subtle differences remain.

Atom interferometers are usually chirped, meaning that the laser light
field is adjusted to the freely falling atom, in order to optimize the reflection rate.
Speaking in the language of dynamical diffraction, the Bragg condition is always fully satisfied.
This is not the case in the neutron interferometer where the lattice planes are in rest in the laboratory system.
If the Bragg condition is fulfilled at the beam splitter, this is no longer the case at the mirror and the analyzer crystal, since the neutrons are
accelerated by gravity between the crystals. Fortunately, the gravitational momentum change is small enough so that the neutrons are still accepted by
the Bragg crystal, but the Bragg condition is no longer exactly fulfilled.

In this near-Bragg regime the two types of interferometer clearly differ. The atoms gain or lose momentum only in the direction of the laser.
The momentum perpendicular to the laser is
not affected and the energy of the atoms is not conserved. On the contrary, the energy of the neutrons is conserved because the underlying scattering process is fully elastic.
Since the momentum transfer perpendicular to the lattice planes is constant ($\hbar H$) also the longitudinal momentum
(tangential to the lattice planes) changes in the near-Bragg case \cite{rauch-werner}. Then the reflection is no longer specular, as shown in detail in the next section.
This feature does not only follow from the dynamical diffraction theory but also from the alternative method of rigorous coupled wave analysis, which
is used to describe diffraction on sinusoidal phase gratings \cite{fally2012, prijatelj2013}.
In an intuitive picture one could imagine that the neutron enters and leaves the crystal at positions with different potential height.
Thereby it loses and gains momentum of different amount in the direction perpendicular to the crystal surface. In the atom case
there is no potential barrier to cross.

In the following calculation we neglect to good approximation the influence of gravity within the crystal, and assume that the neutrons
are only accelerated between the crystals. Also the authors of the original article neglect the crystal thickness and describe the reflection as
an instantaneous process. This is justified since the interaction time within the crystals is much shorter than the flight time between the crystals.
We could also use the accurate solution of Bragg diffraction under gravity \cite{Werner1980, Bonse1984, Werner1988} but it does not change the
basic result of the near-Bragg regime that we want to point out here.

\begin{figure}
\begin{center}\includegraphics[scale=0.75]{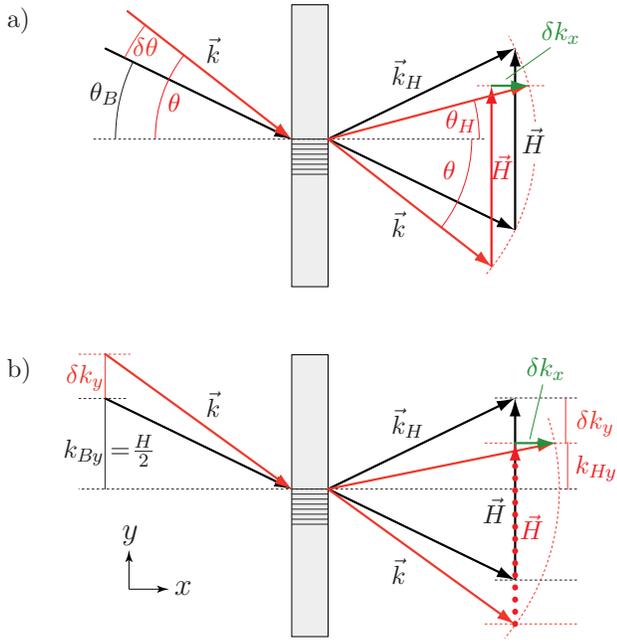}\end{center}
\caption{\label{figlauereflection} Laue reflection for exact Bragg condition (black), Bragg violation by $\delta\theta$ and fixed wave length (a, red) and
Bragg violation by $\delta k_y$ and fixed $k_x$ (b, red). }
\end{figure}

\begin{figure*}
\begin{center}\includegraphics[scale=0.9]{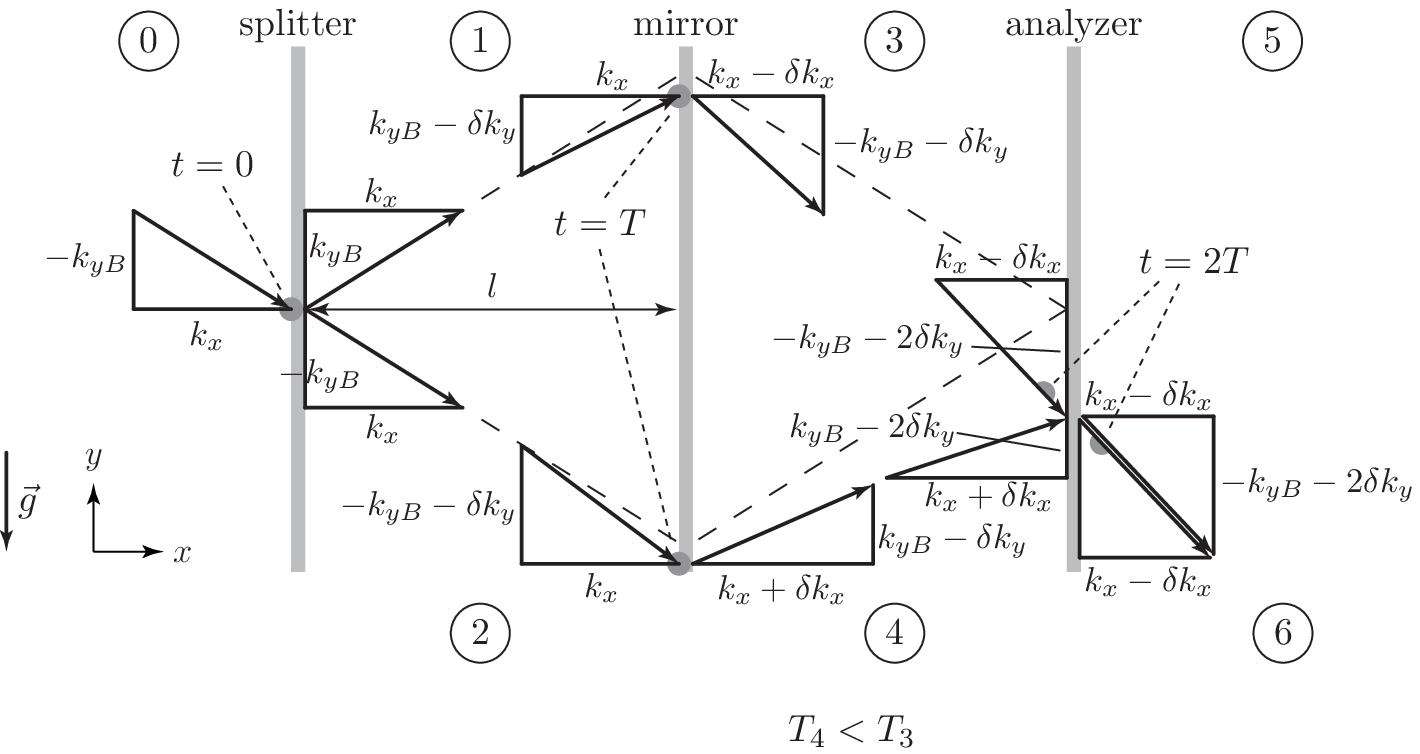}\end{center}
\caption{\label{figifmgrav}
Wave vectors in a neutron interferometer under the influence of gravity.  }
\end{figure*}

\subsection*{Momentum calculation}

Fig. \ref{figlauereflection} illustrates the symmetric Laue case, where the lattice planes of the crystal are perpendicular to the crystal surfaces.  $\vec k$ denotes the wave vector incident on the crystal, $\vec H$ denotes the reciprocal lattice vector.
\begin{align}
  \vec k   &=  \begin{pmatrix}{k_{x}} \\ {k_{y}}\end{pmatrix} = \begin{pmatrix}{k \cos\theta} \\ {-k \sin\theta}\end{pmatrix}  ,   \qquad \vec H=  \begin{pmatrix}{0}\\{H}\end{pmatrix}
\end{align}
In order to calculate the reflected wave vector $\vec k_H$ we don't have to go into
all details of dynamical diffraction. It is sufficient to know that the $k_y$ component, being tangential to the surface,
is conserved when the neutron enters and leaves the crystal, and that inside the crystal the $k_y$ component, which is perpendicular to the lattice planes, can change only by $H$.
It then follows that
\begin{align}
  \vec k_H &=  \begin{pmatrix}{k_{Hx}} \\ {k_{Hy}}\end{pmatrix} = \begin{pmatrix}{\sqrt{k^2-k_{Hy}^2} } \\ {k_y+H}\end{pmatrix}
\end{align}
where the $k_{Hx}$ component follows from energy conservation.

If the Bragg condition is exactly fulfilled (Fig. \ref{figlauereflection} black drawing) we denote the wave vector by $\vec k_B$ and the incident angle by $\theta_B$, and $|k_{By}| = k_B \sin\theta_B = H/2$. Then the reflected wave vector reads $\vec k_H = \vec k_B + \vec H$. The full vertical momentum in inverted, the reflection is specular ($\theta_H = \theta$)
and the process looks like a hard wall reflection. This holds for any $k_x$ as long as $k_y=k_{By}$.

If the Bragg condition is violated by $\delta\theta$ at constant $|\vec k|$ (Fig. \ref{figlauereflection}a, red drawing), the reflected vector is obtained by first adding $\vec H$ to $\vec k$ and then correcting the vector perpendicular to the crystal surface such that the modulus of $\vec k$ is conserved. With a first order approximation for small $\delta\theta$, which is perfectly sufficient for thermal neutrons, incident and reflected angles read
\begin{align}
  \theta   & =\theta_B + \delta\theta \\
  \theta_H & \approx\theta_B - \delta\theta.
\end{align}
The Bragg deviation angle $\delta \theta$ of the reflected beam is inverted,
provided that the angles are defined as modulus of the angle between beam and lattice planes. If $\delta\theta$ was defined as rotation in a global coordinate system, we would conclude that $\delta \theta$ is conserved.

If the Bragg condition is violated by $\delta k_y$ at constant $k_x$ (Fig. \ref{figlauereflection}b, red drawing), the reflected wave vector is obtained by the same considerations and
\begin{align}
  k_y    & = k_{By} + \delta k_y \\
  k_{Hy} & = k_{By} + H_y + \delta k_y = -k_{By} + \delta k_y.
\end{align}
Only $k_{By}$ is inverted by Laue reflection. Any additional $\delta k_y$ component is conserved. However, the longitudinal component $k_x$ is changed by
\begin{align}
  \delta k_x &= k_{Hx} - k_x = \sqrt{k^2-k_{Hy}^2 } - k_{x}\\
    &= k_x \left( \sqrt{1+\frac{2\, \delta k_y \,H}{k_x^2}} -1 \right) \\
    &\approx  \delta k_y \, H/k_x
\end{align}

\subsection*{Momentums in the COW setup}

Fig. \ref{figifmgrav} shows the resulting wave vectors in the whole interferometer loop of the COW setup.
We assume Newtonian gravity and the geometry used in the original paper where the lattice planes of the interferometer are horizontal.
The neutron beam falls onto the first interferometer slab under exact Bragg condition. Between the first and second crystal slab
it accumulates a vertical momentum change of $\delta k_y \hbar= g \,T\!/m$
where $T = lm / (\hbar k_x)$ denotes the flight time and $l$ the distance between the slabs.
At the mirror slab the Bragg condition is violated by $\delta k_y$ which changes the horizontal component by $\pm\delta k_x$ behind the mirror slab.
When the neutron arrives at the analyzer slab another $\delta k_y$ has been accumulated (assuming to lowest order that $T_3\approx T_4 \approx T$)
and the total vertical momentum change reads $2 \delta k_y$. Consequently, the beam that is reflected by the analyzer slab changes its horizontal momentum again by $2\delta k_x$
while the transmitted beam is unchanged. Thus both beam paths end up with the same final wave vector. This is important, otherwise the beams would carry which-way information and
no interference was possible. 

Nevertheless the horizontal momentum transfer changes the flight times. While the flight times in sections 1 and 2 are equal ($T_1=T_2=T$) the flight times behind the mirror slab
differ due to the different horizontal momentum, $T_3>T$, $T_4<T$. This has no impact on the phase shift but defocuses the interferometer and reduces the interference contrast.

Let's look at the numbers for a typical neutron interferometry setup with a 220 silicon interferometer, $\theta_B=30\degree$ and $\lambda=1.9\text{\AA}$.
Bragg diffraction occurs if $k_y$ lies within the Bragg acceptance distribution which has a Lorentzian envelope and the width $\sigma_{ky}/k_y \approx 5\cdot 10^{-6}$.
Assuming $l=5$cm the flight time between two crystal slabs amounts to $T=28$ms and
the gravitational momentum change is  $\delta k_y/k_y=2.7\cdot 10^{-7}$. This is still very well accepted by the mirror slab.
Behind the mirror the horizontal momentum has changed by $\delta k_x/k_x=1.8\cdot 10^{-7}$ and the flight time $T_{3,4}$ in section 3 and 4 changes by
$(T_{3,4}-T)/T=\pm1.8\cdot 10^{-7}$. Behind the analyzer slab the final momentums coming from section 3 and 4 differ by $(k_{y3} - k_{y4})/k_y = 9.5\cdot 10^{-14}$ which
can be neglected.

\subsection*{Conclusion}

We conclude that the flight paths in a neutron interferometer in gravity are to first order equivalent to that of an atom interferometer.
Higher order differences remain for near-Bragg beam components which always arise in the COW setup.

\bibliography{cow_comment}

\begin{thebibliography}{11}%
\makeatletter
\providecommand \@ifxundefined [1]{%
 \@ifx{#1\undefined}
}%
\providecommand \@ifnum [1]{%
 \ifnum #1\expandafter \@firstoftwo
 \else \expandafter \@secondoftwo
 \fi
}%
\providecommand \@ifx [1]{%
 \ifx #1\expandafter \@firstoftwo
 \else \expandafter \@secondoftwo
 \fi
}%
%
\providecommand \enquote  [1]{``#1''}%
\providecommand \bibnamefont  [1]{#1}%
\providecommand \bibfnamefont [1]{#1}%
\providecommand \citenamefont [1]{#1}%
\providecommand \href@noop [0]{\@secondoftwo}%
\providecommand \href [0]{\begingroup \@sanitize@url \@href}%
\providecommand \@href[1]{\@@startlink{#1}\@@href}%
\providecommand \@@href[1]{\endgroup#1\@@endlink}%
\providecommand \@sanitize@url [0]{\catcode `\\12\catcode `\$12\catcode
  `\&12\catcode `\#12\catcode `\^12\catcode `\_12\catcode `\%12\relax}%
\providecommand \@@startlink[1]{}%
\providecommand \@@endlink[0]{}%
\providecommand \url  [0]{\begingroup\@sanitize@url \@url }%
\providecommand \@url [1]{\endgroup\@href {#1}{\urlprefix }}%
\providecommand \urlprefix  [0]{URL }%
%
%
\providecommand \selectlanguage [0]{\@gobble}%
\providecommand \bibinfo  [0]{\@secondoftwo}%
\providecommand \bibfield  [0]{\@secondoftwo}%
%
\providecommand \BibitemOpen [0]{}%
%
%
%
\providecommand \BibitemShut  [1]{\csname bibitem#1\endcsname}%
\let\auto@bib@innerbib\@empty
\bibitem [{\citenamefont {Greenberger}\ \emph {et~al.}(2012)\citenamefont
  {Greenberger}, \citenamefont {Schleich},\ and\ \citenamefont
  {Rasel}}]{greenberger2012}%
  \BibitemOpen
  \bibfield  {author} {\bibinfo {author} {\bibfnamefont {Daniel~M.}\
  \bibnamefont {Greenberger}}, \bibinfo {author} {\bibfnamefont {Wolfgang~P.}\
  \bibnamefont {Schleich}}, \ and\ \bibinfo {author} {\bibfnamefont {Ernst~M.}\
  \bibnamefont {Rasel}},\ }\bibfield  {title} {\enquote {\bibinfo {title}
  {Relativistic effects in atom and neutron interferometry and the differences
  between them},}\ }\href@noop {} {\bibfield  {journal} {\bibinfo  {journal}
  {Physical Review A}\ }\textbf {\bibinfo {volume} {86}},\ \bibinfo {pages}
  {063622} (\bibinfo {year} {2012})}\BibitemShut {NoStop}%
\bibitem [{\citenamefont {Schleich}\ \emph {et~al.}(2013)\citenamefont
  {Schleich}, \citenamefont {Greenberger},\ and\ \citenamefont
  {Rasel}}]{schleich2013}%
  \BibitemOpen
  \bibfield  {author} {\bibinfo {author} {\bibfnamefont {Wolfgang~P.}\
  \bibnamefont {Schleich}}, \bibinfo {author} {\bibfnamefont {Daniel~M.}\
  \bibnamefont {Greenberger}}, \ and\ \bibinfo {author} {\bibfnamefont
  {Ernst~M.}\ \bibnamefont {Rasel}},\ }\bibfield  {title} {\enquote {\bibinfo
  {title} {{A representation-free description of the Kasevich-Chu
  interferometer: a resolution of the redshift controversy}},}\ }\href@noop {}
  {\bibfield  {journal} {\bibinfo  {journal} {{New Journal of Physics}}\
  }\textbf {\bibinfo {volume} {15}},\ \bibinfo {pages} {013007} (\bibinfo
  {year} {2013})}\BibitemShut {NoStop}%
\bibitem [{\citenamefont {Batterman}\ and\ \citenamefont
  {Cole}(1964)}]{batterman1964}%
  \BibitemOpen
  \bibfield  {author} {\bibinfo {author} {\bibfnamefont {B.~W.}\ \bibnamefont
  {Batterman}}\ and\ \bibinfo {author} {\bibfnamefont {H.}~\bibnamefont
  {Cole}},\ }\bibfield  {title} {\enquote {\bibinfo {title} {Dynamical
  diffraction of x rays by perfect crystals},}\ }\href@noop {} {\bibfield
  {journal} {\bibinfo  {journal} {Review of Modern Physics}\ }\textbf {\bibinfo
  {volume} {36}},\ \bibinfo {pages} {681--717} (\bibinfo {year}
  {1964})}\BibitemShut {NoStop}%
\bibitem [{\citenamefont {Rauch}\ and\ \citenamefont
  {Werner}(2000)}]{rauch-werner}%
  \BibitemOpen
  \bibfield  {author} {\bibinfo {author} {\bibfnamefont {Helmut}\ \bibnamefont
  {Rauch}}\ and\ \bibinfo {author} {\bibfnamefont {Samuel~A.}\ \bibnamefont
  {Werner}},\ }\href@noop {} {\emph {\bibinfo {title} {Neutron
  Interferometry}}}\ (\bibinfo  {publisher} {Clarendon Press},\ \bibinfo
  {address} {Oxford},\ \bibinfo {year} {2000})\BibitemShut {NoStop}%
\bibitem [{\citenamefont {Petrascheck}(1976)}]{petrascheck1976}%
  \BibitemOpen
  \bibfield  {author} {\bibinfo {author} {\bibfnamefont {Dietmar}\ \bibnamefont
  {Petrascheck}},\ }\bibfield  {title} {\enquote {\bibinfo {title} {Theorie
  eines neutroneninterferometers},}\ }\href@noop {} {\bibfield  {journal}
  {\bibinfo  {journal} {Acta Physica Austriaca}\ }\textbf {\bibinfo {volume}
  {45}},\ \bibinfo {pages} {217--231} (\bibinfo {year} {1976})}\BibitemShut
  {NoStop}%
\bibitem [{\citenamefont {Lemmel}(2007)}]{lemmel2007}%
  \BibitemOpen
  \bibfield  {author} {\bibinfo {author} {\bibfnamefont {Hartmut}\ \bibnamefont
  {Lemmel}},\ }\bibfield  {title} {\enquote {\bibinfo {title} {Dynamical
  diffraction of neutrons and transition from beam splitter to phase shifter
  case},}\ }\href@noop {} {\bibfield  {journal} {\bibinfo  {journal} {Physical
  Review B}\ }\textbf {\bibinfo {volume} {76}},\ \bibinfo {pages} {144305}
  (\bibinfo {year} {2007})}\BibitemShut {NoStop}%
\bibitem [{\citenamefont {Fally}\ \emph {et~al.}(2012)\citenamefont {Fally},
  \citenamefont {Klepp},\ and\ \citenamefont {Tomita}}]{fally2012}%
  \BibitemOpen
  \bibfield  {author} {\bibinfo {author} {\bibfnamefont {Martin}\ \bibnamefont
  {Fally}}, \bibinfo {author} {\bibfnamefont {J\"urgen}\ \bibnamefont {Klepp}},
  \ and\ \bibinfo {author} {\bibfnamefont {Yasuo}\ \bibnamefont {Tomita}},\
  }\bibfield  {title} {\enquote {\bibinfo {title} {An experimental study on the
  validity of diffraction theories for off-bragg replay of volume holographic
  gratings},}\ }\href@noop {} {\bibfield  {journal} {\bibinfo  {journal}
  {Applied Physics B}\ }\textbf {\bibinfo {volume} {108}},\ \bibinfo {pages}
  {89--96} (\bibinfo {year} {2012})}\BibitemShut {NoStop}%
\bibitem [{\citenamefont {Prijatelj}\ \emph {et~al.}(2013)\citenamefont
  {Prijatelj}, \citenamefont {Klepp}, \citenamefont {Tomita},\ and\
  \citenamefont {Fally}}]{prijatelj2013}%
  \BibitemOpen
  \bibfield  {author} {\bibinfo {author} {\bibfnamefont {Matej}\ \bibnamefont
  {Prijatelj}}, \bibinfo {author} {\bibfnamefont {J\"urgen}\ \bibnamefont
  {Klepp}}, \bibinfo {author} {\bibfnamefont {Yasuo}\ \bibnamefont {Tomita}}, \
  and\ \bibinfo {author} {\bibfnamefont {Martin}\ \bibnamefont {Fally}},\
  }\bibfield  {title} {\enquote {\bibinfo {title} {Far-off-bragg reconstruction
  of volume holographic gratings: a comparison of experiment and theories},}\
  }\href@noop {} {\bibfield  {journal} {\bibinfo  {journal} {Physical Review
  A}\ }\textbf {\bibinfo {volume} {87}},\ \bibinfo {pages} {063810} (\bibinfo
  {year} {2013})}\BibitemShut {NoStop}%
\bibitem [{\citenamefont {Werner}(1980)}]{Werner1980}%
  \BibitemOpen
  \bibfield  {author} {\bibinfo {author} {\bibfnamefont {Samual~A.}\
  \bibnamefont {Werner}},\ }\bibfield  {title} {\enquote {\bibinfo {title}
  {Gravitational and magnetic field effects on the dynamical diffraction of
  neutrons},}\ }\href@noop {} {\bibfield  {journal} {\bibinfo  {journal}
  {Physical Review B}\ }\textbf {\bibinfo {volume} {21}},\ \bibinfo {pages}
  {1774--1789} (\bibinfo {year} {1980})}\BibitemShut {NoStop}%
\bibitem [{\citenamefont {Bonse}\ and\ \citenamefont
  {Wroblewski}(1984)}]{Bonse1984}%
  \BibitemOpen
  \bibfield  {author} {\bibinfo {author} {\bibfnamefont {Ulrich}\ \bibnamefont
  {Bonse}}\ and\ \bibinfo {author} {\bibfnamefont {Thomas}\ \bibnamefont
  {Wroblewski}},\ }\bibfield  {title} {\enquote {\bibinfo {title} {Dynamical
  diffraction effects in noninertial neutron interferometry},}\ }\href@noop {}
  {\bibfield  {journal} {\bibinfo  {journal} {Phys. Rev. D}\ }\textbf {\bibinfo
  {volume} {30}},\ \bibinfo {pages} {1214--1217} (\bibinfo {year}
  {1984})}\BibitemShut {NoStop}%
\bibitem [{\citenamefont {Werner}\ \emph {et~al.}(1988)\citenamefont {Werner},
  \citenamefont {Kaiser}, \citenamefont {Arif},\ and\ \citenamefont
  {Clothier}}]{Werner1988}%
  \BibitemOpen
  \bibfield  {author} {\bibinfo {author} {\bibfnamefont {S.A.}\ \bibnamefont
  {Werner}}, \bibinfo {author} {\bibfnamefont {H.}~\bibnamefont {Kaiser}},
  \bibinfo {author} {\bibfnamefont {M.}~\bibnamefont {Arif}}, \ and\ \bibinfo
  {author} {\bibfnamefont {R.}~\bibnamefont {Clothier}},\ }\bibfield  {title}
  {\enquote {\bibinfo {title} {Neutron interference induced by gravity: New
  results and interpretations},}\ }\href@noop {} {\bibfield  {journal}
  {\bibinfo  {journal} {Physica B}\ }\textbf {\bibinfo {volume} {151}},\
  \bibinfo {pages} {22 -- 35} (\bibinfo {year} {1988})}\BibitemShut {NoStop}%
\end{thebibliography}

\end{document}